\documentclass[
    ,final            
 ,numberedheadings 
  ]
  {aipproc}

\layoutstyle{8x11single}

\begin{document}

\title{High-energy direct reactions with exotic nuclei
and low-energy nuclear astrophysics}


\author{G.Baur}{
  address={Institut f\"{u}r Kernphysik, Forschungszentrum J\"{u}lich, 
    D-52425 J\"{u}lich, Germany}
}

\author{S.Typel}{
  address={Gesellschaft f\"{u}r Schwerionenforschung mbH (GSI),
           Planckstra\ss{}e 1, D-64291 Darmstadt, Germany}
}

\begin{abstract}
Indirect methods in nuclear astrophysics 
are discussed. Recent work on Coulomb dissociation
and an effective-range theory of low-lying electromagnetic 
strength of halo nuclei is presented.
Coulomb dissociation of a halo nucleus bound by a zero-range 
potential is proposed  as a homework problem
(for further references see G.~Baur and S.~Typel, nucl-th/0504068).
It is pointed out that the Trojan-Horse method 
(G.~Baur, F.~R\"{o}sel, D.~Trautmann and R.~Shyam, Phys. Rep. 111 (1984) 333) 
is a suitable tool to investigate subthreshold resonances.
\end{abstract}

\maketitle


\section{Introduction and Overview}


With the exotic beam facilities all over 
the world - and more are to come - direct reaction theories 
are experiencing a renaissance. 
We give a minireview of indirect methods 
for nuclear astrophysics reporting on recent work on 
Coulomb dissociation of halo nuclei \cite{tybaprl,tyba04}
and on transfer reactions to bound and scattering states.
The chemical evolution 
of the universe and the role of radioactive beams has recently been
reviewed in \cite{brad}.
In \cite{sam} it is remarked that high energy (radioactive)
beams are a valuable tool to obtain information on low
energy nuclear reactions of astrophysical importance.
We discuss Coulomb dissociation \cite{bbr} and the Trojan Horse
method \cite{Bau76,baur} as examples.

Coulomb dissociation of a neutron halo nucleus 
in the limit of a zero-range neutron-core interaction
in the Coulomb field of a target nucleus can be 
studied in various limits of the parameter space
and rather simple analytical solutions can be found. 
We propose to solve the scattering problem for 
this model Hamiltonian by means of 
the various advanced numerical methods that are available nowadays.
In this way their range of applicability can be studied
by comparison to the analytical benchmark solutions, for 
work in this direction see \cite{banerj}. 

The Trojan-Horse Method \cite{tyba02,fus03} is a particular case
of transfer reactions to the continuum under quasi-free scattering
conditions.
Special attention is paid to the 
transition from reactions to bound and unbound states
and the role of subthreshold resonances.
Since the binding energies of nuclei close to the drip
line tend to be small, this is expected to be 
an important general feature in  
exotic nuclei. 

\section{Effective Range Theory of Halo Nuclei}

At low energies the effect of the nuclear potential is 
conveniently described by the effective-range expansion
\cite{Bet49}.
An effective-range approach for the electromagnetic
strength distribution in neutron halo 
nuclei was introduced in \cite{tybaprl} and
applied to the single neutron halo nucleus ${}^{11}$Be.
Recently, the same method was applied to the description of
electromagnetic dipole strength in ${}^{23}$O
\cite{Noc04}.
A systematic study sheds 
light on the sensitivity of the electromagnetic 
strength distribution to the interaction in the continuum.
We expose the dependence on the binding energy of the nucleon
and on the angular
momentum quantum numbers. Our approach extends the familiar textbook
case of the deuteron,
that can be considered as the
prime example of a halo nucleus, to arbitrary nucleon+core systems,
for related work see \cite{kala,besu,be}.
We also investigate in detail the square-well potential model.
It has  great merits: it can be solved analytically,
it shows the main characteristic features
and it leads to rather simple and transparent formulae.
As far as we know, some of these formulae have not been published before.
These explicit results can be compared to our general
theory for low energies (effective-range approach) and also 
to more realistic Woods-Saxon models. Due to shape independence,
the results of these various approaches will not differ for
low energies. It will be interesting to delineate
the range of validity of the simple models.  

Our effective-range approach is closely related to effective field theories
that are nowadays used for the description of 
the nucleon-nucleon system and halo nuclei
\cite{Ber02}. 
The characteristic low-energy
parameters are linked to QCD in systematic expansions.
Similar methods are also used in 
the study of exotic atoms ($\pi^-A$, $\pi^+\pi^-$, $\pi^-p$, \dots) 
in terms of effective-range parameters.
The close relation of effective field theory to the effective-range
approach for hadronic atoms was discussed in Ref.\ \cite{Hol99}.

\begin{figure}
\includegraphics[height=.3\textheight]{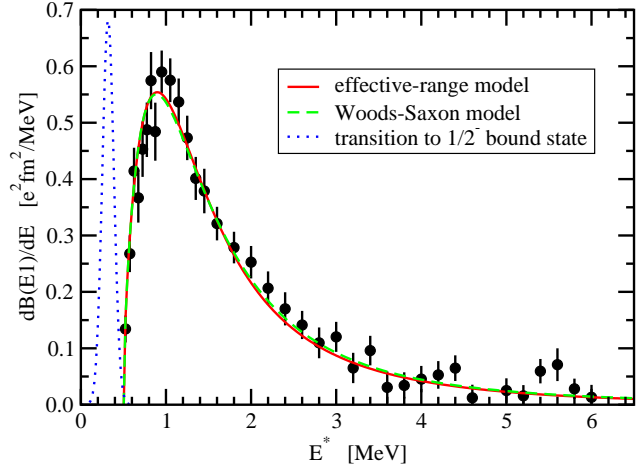}
\caption{\label{fig:1} Reduced probability for
dipole transitions as a function
of the excitation energy $E^{\ast}=E+S_{n}$
in comparison to experimental data extracted from
Coulomb dissociation of ${}^{11}$Be \cite{palit}.}
\end{figure}

In Fig.~\ref{fig:1} we show the application of the method to 
the electromagnetic dipole strength in $^{11}$Be. 
The reduced transition probability was deduced from high-energy
${}^{11}$Be Coulomb dissociation at GSI \cite{palit}.
Using a cutoff radius of $R=2.78$~fm and 
an inverse bound-state decay length of
$q=0.1486$~fm${}^{-1}$ as input parameters we extract
an ANC of $C_{0}=0.724(8)$~fm${}^{-1/2}$ 
from the fit to the experimental data. The ANC
can be converted to a spectroscopic factor of $C^{2}S=0.704(15)$
that is consistent with results from other methods.
In the lowest order of 
the effective-range expansion the phase shift 
$\delta_{l}^{j}$
in the partial wave with orbital angular momentum $l$ and
total angular momentum $j$
is written as $\tan \delta_{l}^{j}= -(x c_{l}^{j}\gamma)^{2l+1}$, 
where $\gamma=qR=0.4132<1$ is the halo expansion parameter and $x=k/q
=\sqrt{E/S_{n}}$ with the neutron separation energy $S_{n}$. 
The effective range term $\frac{1}{2} r_l k^2$ term can be neglected,
since it leads to a contribution with an extra $\gamma^2$ factor
which is small in the halo nucleus limit $\gamma \to 0$
(at least in the case where the scattering length is of natural
order).
The parameter  $c_{l}^{j}$ corresponds to the scattering
length $a_{l}^{j} = (c_{l}^{j}R)^{2l+1}$. We obtain
$c^{3/2}_{1}=-0.41(86,-20)$ and $c^{1/2}_{1}=2.77(13,-14)$. The latter 
is unnaturally large because of the existence of a bound $\frac{1}{2}^{-}$
state close to the neutron breakup threshold in ${}^{11}$Be.

 The connection of the scattering length $a_l$ and the bound state
parameter q for $l>0$ is given by
$a_l=\frac{2 (2l-1) R^{2l-1}}{q^2(2l+1)!!(2l-1)!!}$.
This is a generalization of the
well-known relation $a_0=1/q$ for $l=0$ in a square well model,where
R denotes the range of the potential.
The $p_{1/2}$ channel in $^{11}$Be is an example for the 
influence of a halo state on the continuum. 
The binding energy
of this state is given by 184 keV, which corresponds to $q=0.094$~fm. 
With $R=2.78$~fm one has $\gamma^2=0.068$.
For $l=1$ one has $a_1=\frac{2R^3}{3\gamma^2}=210$~ fm$^3$ 
which translates into $c_1= (a_1/R^3)^{1/3}=2.14$. This compares
favourably with the fit value given in table 1 of \cite{tybaprl}: 
$2.77(13,-14)$.

For a further discussion we refer to \cite{tybaprl}.

\section{A solvable  model for Coulomb Dissociation of neutron halo nuclei}

We consider a three-body system consisting of a neutron n, a core c
and an (infinitely heavy) target nucleus with charge Ze. 
The Hamiltonian is given by
\begin{equation}
H=T_r+T_{r_c}+V_{cZ}(r_c)+ V_{nc}(r)
\end{equation}
where $T=T_r+ T_{r_c}$ is the kinetic energy of the system.
The Coulomb interaction between the core and the target
is given by $V_{cZ}=Z Z_{c} e^2/r_{c}$ 
and $V_{nc}$ is a zero-range interaction between 
$c$ and $n$. The s-wave bound state of the $a=(c+n)$ system
is given by the wave function $\Phi_0=\sqrt{q/(2\pi)}
\exp{(-qr)}/r$, where $q$ is related to the binding energy
$E_{b}$ by $ E_{b} = \hbar^{2}q^{2}/(2\mu)$ and
the reduced mass of the $c+n$ system
is denoted by $\mu$. 
This system can be studied analytically in various approximations.
It can serve as a benchmark for the comparison of various analytical
as well as numerical approaches. 
We refer to \cite{ppnp} 
(see especially Ch.~4 there) for details. 

The kinematics of the breakup process is given by 
$ \vec{q}_{a} \rightarrow \vec{q}_{cm} + \vec{q}_{rel}$ where
$\vec{q}_{cm}$ and $\vec{q}_{rel}$ are directly 
related to $\vec{q}_{c}$ and $\vec{q}_{n}$, respectively.
Analytic results are known for the plane-wave limit,
the Coulomb-wave Born approximation (CWBA,
``Bremsstrahlung integral'') and the adiabatic approximation \cite{jeff}).
A first derivation  of the ``Bremsstrahlung formula'' was given by 
Landau and Lifshitz \cite{ll}, it was improved by 
Breit in \cite{breit}; an early review is given in \cite{bata}.
It was first applied to heavy ions ($^{11}$Be) in \cite{sbg}.

In the plane-wave limit the result does not depend on 
$q_{a}$ itself but only on the ``Coulomb push'' 
$\vec{q}_{coul}=\vec{q}_{a} - \vec{q}_{cm}$.
In the semiclassical high energy straight-line 
and electric dipole limit, first and second order
analytical results are available, as well as for the sudden limit.
E.g., in the straight-line dipole approximation 
a shape parameter $x=k/q$ and a strength parameter 
$y=m_{n}\eta/[(m_{n}+m_{c})b q]$ determine the 
breakup probability (in the sudden limit). The impact parameter
is denoted by $b$ and the Coulomb parameter is
$\eta=Z Z_c e^2/(\hbar v)$. 
In \cite{tyba01} it was found that the breakup probability
is given in leading order by 
\begin{equation}
\frac{dP_{LO}}{dk}=\frac{16}{3\pi q}
y^{2} \frac{x^{4}}{(1+x^{2})^{4}}
\end{equation}
and in next-to-leading order by 
\begin{equation}
\frac{dP_{NLO}}{dk}=\frac{16}{3\pi q}
 y^{4} \frac{x^{2}(5-55x^{2}+28x^{4})}{15(1+x^{2})^{6}}.
\end{equation}
Depending on the parameter $x$, the latter contribution leads 
to an enhancement or a reduction
of the breakup probability as compared to the leading-order result.
Another important scaling parameter,
in addition to $x$ and $y$, is $\xi=\omega b/v$,
where $\hbar \omega$ is the excitation energy of the $(c+n)$ system.
In the sudden approximation we have $\xi=0$
and there is an analytical solution \cite{tybanuc}.

The dependence of the post-acceleration effect on the beam energy
was  studied in  post-form CWBA
in  \cite{banerj}. Postacceleration is very important for low
beam energies and tends to diminish with high energies,
see especially Sect. 4.2 of \cite{ppnp}.
This may pose a problem for the 
CDCC approach at low beam energies.
The choice of the Jacobi coordinates to represent the CDCC
basis is discussed in \cite{nun}.

\section{Recent experimental results and astrophysical applications}

The status of Coulomb dissociation has been reviewed
until about 2003 in \cite{ppnp}. In addition to the
general theoretical framework, this review also contains a discussion
of the experimental results along with their astrophysical significance.  
The last years saw progress for various cases.
This is a very brief summary of the experimental results.

The $^7$Be($p,\gamma$)$^8$B reaction plays a key role
in the determination of the solar neutrino flux.
$^8$B Coulomb dissociation experiments have been carried out
over decades at RIKEN, MSU and GSI, with increasing beam energies.
A careful analysis of the 
GSI $^8$B experiment 'GSI-2' is now  avialable \cite{schue},
see also \cite{ntaa}. 
Particular emphasis was placed on the angular correlations of the
breakup particles.These correlations demonstrate clearly
that $E1$ multipolarity dominates. The deduced astrophysical $S_{17}$
factor shows good agreement with most recent direct $^7$Be($p,\gamma$)$^8$B
measurements. 
High beam energies help to reduce higher order effects.
The equivalent photon spectrum weights the $E2$ contribution 
more than the $E1$ contribution. This effect diminishes with increasing
beam energy. Thus the high beam energy of 254~A~MeV used at GSI
helps to reduce the importance of the $E2$ contribution. 

The origin of $^6$Li in the early universe is an interesting topic at present
\cite{reeves}. $^6$Li is produced via the $\alpha+d \rightarrow ^6$Li$+
\gamma$ radiative capture reaction in big bang nucleosynthesis. 
The S factor of this radiative capture reaction can be determined
in a $^6$Li Coulomb breakup experiment. (The fragility of $^6$Li is 
due to the large cross section of the
$^6$Li($p,\alpha$)$^3$He reaction, see Ch. 5.3 below).
The results of the $^6$Li 
$\rightarrow \alpha +d$ Coulomb dissociation experiment
at $E_{^6Li}=$150~A~MeV at the KaoS spectrometer at GSI
have recently been presented \cite{Ham05}.
Compared to the pilot experiment with $E_{^6Li}=26$~A~MeV \cite{kiener},
lower
$\alpha-d$ relative energies down to 50 keV could be reached with
rather small error bars. Again, the high beam energy is very useful.  

Experimental results for the Coulomb breakup of psd-shell
neutron rich nuclei from GSI were presented in \cite{ushasi}.
 There is valuable spectroscopic 
information on various isotopes. 
The observed electromagnetic $E1$ strength above the one-neutron 
threshold of neutron-rich C, Be, B and O isotopes is explained 
by a non-resonant transition of a neutron into the continuum. 
The effective-range theory of halo nuclei given in Ch.\ 2 is
well suited to describe these effects. 
The neutron capture cross section for the $^{14}$C($n,\gamma$)$^{15}$C
reaction, which is of astrophysical relevance, is also deduced.
The discrepancy between the 
Coulomb dissociation eperiments at GSI \cite{ushasi,plb03}
and MSU as well as with the $(n,\gamma)$-
capture results from Karlsruhe still persists.

Recently the E1-strength above neutron treshold 
in the neutron-rich $^{130}$Sn and $^{132}$Sn isotopes
was investigated in a Coulomb dissociation experiment 
at GSI \cite{au}. Low lying strength was found ('pygmy resonance').
The authors conclude that 'this E1 strength arises
from oscilllations of the excess neutrons but it is
under debate to what extent a collective mode is formed'. 

The explosive hydrogen burning process synthesizes
nuclei in hot, dense and hydrogen-rich stellar
environments. Proton capture processes are important to
understand this rp-process. 
Progress has been achieved in the Coulomb dissociation 
of p-rich nuclei which is an indirect way to 
study these capture processes.
There are examples of Coulomb dissociation of
proton rich nuclei which allow to deduce
S factors for $(p,\gamma)$ reactions.
T.Gomi et al. 
\cite{gomi,gomineu} performed a Coulomb dissociation
experiment on $^{23}$Al in order to study the stellar
$^{22}$Mg($p,\gamma$)$^{23}$Al reaction.
They determine the radiative width $\Gamma_\gamma$
of the first excited state in $^{23}$Al by measuring
the $^{23}$Al$ \rightarrow ^{22}$Mg $+p$ Coulomb dissociation.
The Coulomb breakup of $^{27}$P was studied in \cite{toga}
in order to determine 
the S factor for the  $^{26}$Si($p,\gamma$)$^{27}$P reaction. 
The $E2$ $\gamma$-decay width of the first excited state
in $^{27}$P was extracted. Assuming the same $E2/M1$ mixing
ratio as the one for the mirror nucleus $^{27}$Mg
the astrophysically interesting $\gamma$-width can be estimated.
Coulomb dissociation experiments with ${}^{23}$Al and ${}^{27}$P are
also planned at GSI with higher beam energies in the near future.

For heavier nuclei with increasing charge  the continuum contributions
will diminish, due to the increase of the Coulomb barrier.
The S factor will be dominated by resonance contributions that can be 
calculated in the Hauser-Feshbach approach, if the statistical assumptions
are fulfilled.

\section{Transfer Reactions}

Exotic nuclei have low thresholds for particle emission.
It is expected that in transfer reactions one will
often meet a situation where the transferred particle is 
in a state close to the particle threshold. 
In ``normal'' nuclei, the neutron threshold is
around an excitation energy of about 8 MeV, and 
the pure single particle picture is not directly applicable.
Much is known from stripping reactions like $(d,p)$
and thermal neutron scattering, see, e.g., \cite{bomo}.
The single-particle strength
is fragmented over many more complicated compound states. The 
interesting quantity is the strength function.
It is proportional to  $\Gamma/D$
where $\Gamma$ is the width and D the level spacing.
One has  $\Gamma/D \ll 1$, as can be estimated from a 
square well model (see, e.g., \cite{bomo}).

For neutron rich (halo) nuclei the neutron threshold
is much lower, of the order of one MeV. In this
case the single-particle properties
are dominant and the ideas developed in the following can become relevant,
see also \cite{blan}. The level density is also much lower. 
In normal nuclei the level density at particle threshold is generally
so high that the single-particle structure is very much
dissolved. This can be quite different in exotic nuclei
which can show a very pronounced single-particle 
structure. 

\subsection{Trojan-Horse Method}

A similarity between cross sections for two-body and closely
related three-body reactions under certain kinematical conditions
\cite{Fuc71}
led to the introduction of the Trojan-Horse method 
\cite{Bau76,Bau84,Typ00,tyba02}.
In this indirect approach a two-body reaction
\begin{equation} \label{APreac}
 A + x \to C + c
\end{equation}
that is relevant to nuclear astrophysics is replaced by a reaction
\begin{equation} \label{THreac}
 A + a \to C + c + b
\end{equation}
with three particles in the final state. 
One assumes that the Trojan horse
$a$ is composed predominantly of clusters $x$ and $b$, i.e.\  $a=(x+b)$. 
This reaction can be considered as a special case of a transfer 
reaction to the continuum. It is studied experimentally under quasi-free
scattering conditions, i.e.\ when the momentum transfer to the
spectator $b$ is small. The method was primarily applied to the
extraction of the low-energy cross section of reaction
(\ref{APreac}) that is relevant for astrophysics. However, the method
can also be applied to the study of single-particle states in exotic
nuclei around the particle threshold.

The basic assumptions of the Trojan Horse Method are discussed
in detail in \cite{tyba02}, see especially Section 2 there. 
In view of a recent preprint \cite{muk}
we give here a very short outline of the reasoning
(see also \cite{fus03}).
The method is based on the assumption that the transfer of 
particle $x$ is a direct reaction process, for which the 
well known DWBA description \cite{austern}
is appropriate. In contrast to \cite{muk} no such assumption is needed
for the subprocess $A+x \to c+C$, which is of interest in this context.
Although the post- and prior forms are equivalent, it is
simpler to use the post form DWBA. The equivalence of 
post and prior sum rules for inclusive breakup reactions 
are elucidated in \cite{iav}('IAV'), see also \cite{mun}, 
where details also to the more formal aspects and further references
can be found. The basic approximation of \cite{tyba02} is the 
surface approximation, which allows one to relate the
T-matrix of the process (\ref{THreac}) to the (on-shell) S-matrix elements
of the astrophysical reaction (\ref{APreac}). The surface approximation was
checked numerically in the inclusive breakup formalism 
of IAV in \cite{ki}. It proved to be very well fulfilled in this
example.  This exercise can be performed for every individual case
in a similar manner.  

\subsection{Continuous Transition from Bound to Unbound State
Stripping}

Motivated by this 
we look again at the relation between transfer to 
bound and unbound states. Our notation is as follows:
we have the reaction
\begin{equation}
A+a \rightarrow B+b
\end{equation}
where $a=(b+x)$ and B denotes the final
$B=(A+x)$ system. It can be a bound state
$B$ with binding energy $E_{bind}=-E_{Ax}(>0)$,
the open channel $A+x, \mbox{with }E_{Ax}>0$, or 
another channel $C+c$ of the system $B=(A+x)$.
In particular, the reaction $x+A \rightarrow C+c$ can have
a positive $Q$ value and the energy $E_{Ax}$ can be negative
as well as positive.
As an example we quote the recently studied Trojan horse reaction
$d$+${}^{6}$Li \cite{auro03} applied to the ${}^{6}$Li$(p,\alpha)^{3}$He
two-body reaction (the neutron being the spectator).
In this case there are two charged
particles in the initial state (${}^{6}$Li+$p$).
Another example with a neutral particle $x$
would be ${}^{10}$Be$ + d \rightarrow p
+{}^{11}$Be$ +\gamma$.  
The general question which we want to answer
here is how the two regions $E_{Ax}>0$ and
$E_{Ax}<0$ are related to each other. 
E.g., in Fig.\ 7 of \cite{auro03} the coincidence yield 
is plotted as a function of the ${}^{6}$Li-$p$ relative energy.
It is nonzero at zero relative energy. How does 
the theory \cite{tyba02} (and the experiment)
continue to negative relative energies?
With this method, subtreshold resonances can be 
investigated rather directly.
We treat two cases separately, one where
system $B$ is always in the $(A+x)$ channel,
with a real potential $V_{Ax}$ between $A$ and $x$.
In the other case, there are also other channels
$C+c$, at positive and negative energies $E_{Ax}$.

The cross section is a quantity which only
exists for $E_{Ax}>0$. However, a quantity like the S factor
(or related to it)
can be continued to energies below the threshold.
An instructive example is the modified shape function $\tilde{S}$ in Ch.\ 6
of \cite{tyba04}. In analogy to the astrophysical S factor,
where the Coulomb barrier is taken out,  the
angular momentum barrier is taken out
in the quantity $\tilde{S}$. As can be seen from table 3 or 4
of \cite{tyba04} the quantity $\tilde{S}$ is well defined 
for $x^2<0$, with the characteristic pole at $x^2=-1$,
corresponding to the binding energy of the $(A+x)=B$-system.
 
We refer to \cite{msuproc} for further discussion.

\subsection{Some recent experiments using the Trojan Horse Method 
for  nuclear astrophysics}

It is mainly the Catania group led by Claudio Spitaleri that 
has shown in many examples how the Trojan-horse method can be developed into
a useful tool for nuclear astrophysics. Many 
interesting results have been obtained which we 
summarize very briefly below. An especially interesting aspect is the 
following: 

At sufficiently low energies, electronic screening 
affects the cross section of astrophysical reactions.
For a recent experimental study see \cite{LUNA},
a theoretical analysis is provided in \cite{barker}.
These effects depend on the laboratory environments and can 
also be different from the astrophysical conditions.
Due to the high beam energy in the Trojan Horse
Method, there is no screening of the Coulomb potential by the electron
cloud and one determines the bare nucleus astrophysical S factor.
The knowledge of this bare S factor - derived with
the help of  an indirect method - is useful
in judging the screening effects in the 'direct'
reaction under specific laboratory conditions.
One can also determine the S factor at higher energies
where screening is negligible. For the application of the
Trojan-horse method it is mainly
necessary that the theoretical description yields the correct 
energy dependence, and not necessarily the absolute
value of the cross section. Eventually one has 
to determine - by means of model calculations - the 
astrophysical S factor under the
astrophysical conditions, this is the
quantity one is most interested in.    
For a list of experiments see Sect. 6 and Tab. 1 of \cite{tyba02}
and also the introduction of \cite{spita}.

The two stable isotopes of Li with $A=6,7$, 
are valuable probes for conditions in the early universe. 
While it is safe to say that the 
issue is far from settled at the present times, input from
nuclear astrophysics is certainly important.
The formation of $^6$Li in the early universe proceeds mainly 
by the $\alpha+d$ radiative capture process mentioned in Sect. 4,
the reactions which destroy ${}^{6}$Li have been studied 
using the Trojan-horse method.

The destruction of $^6$Li can proceed by the reaction
$d(^6$Li,$\alpha)\alpha$. This reaction was studied by 
the Trojan Horse reaction $^6$Li($^6$Li,$\alpha \alpha)^4$He,
where one of the  $\alpha$ particles has to be considered
as a spectator ('Trojan Horse') \cite{claudi,aga,gianl}.
The spectator condition, i.e.\ small momentum transfer in the three-body
reaction, is ensured by the quasi-free kinematics.

Another reaction which depletes $^6Li$ is $p(^6$Li,$\alpha)^3$He.
It was studied in \cite{auro03}
by means of the Trojan Horse
reaction  $^2$H($^6$Li,$\alpha {}^3$He)$n$. Coincidence spectra were
measured in a kinematically complete experiment at $E_{^6Li}=25$~MeV.
They show the presence of the quasifree $^6$Li-$p$ process. 
Cross sections for the $^6$Li($p,\alpha)^3$He from $E_{cm}=2.4$~MeV
down to astrophysical energies were extracted. (Actually, 
the experimental results 
extend also to negative $^6$Li-$p$-energies).  

The $^{11}$B($p,\alpha_0)^8$Be 
reaction was studied from 1~MeV down to astrophysical
energies by means of the Trojan-horse method applied to the 
three-body reaction $d(^{11}$B,$\alpha_0 ^8$Be)$n$ \cite{spita}.
This reaction is responsible for the boron destruction in stellar environments.

The reaction $^3$He($d,p)^4$He was studied in \cite{marco}
by means of the $^6$Li($^3$He,$p\alpha)^4$He
three-body reaction. The bare astrophysical $S(E)$
factor was deduced. This allowed 
an independent estimate of the screening potential, confirming
the discrepancy with the adiabatic limit. The reaction $^3$He($d,p)^4$He
is important for primordial nucleosynthesis \cite{barker}. 

One may also envisage applications of the Trojan-horse method with exotic
beams. An unstable (exotic) projectile hits a Trojan-horse target
allowing to study specific reactions on exotic nuclei that are unaccessible
in direct experiments. We mention the $d(^{56}$Ni,$ p)^{57}$Ni
reaction studied in inverse kinematics in 1998 \cite{ernst}. In this
paper, stripping to bound states was studied.
In the meantime, more (d,p) transfer reactions  were 
studied in inverse kinematics \cite{jones,wuos,thomas}.
 An extension to 
stripping into the continuum would be of interest for this 
and other kinds of reactions. 

\section{Conclusion}

While the foundations of direct reaction theory
have been laid several decades ago, the new possibilites which have 
opened up with the rare isotope beams 
are an invitation to revisit this field. The general frame 
is set by nonrelativistic many-body quantum scattering theory,
however, the increasing level of precision demands
a good understanding of relativistic effects notably in 
intermediate-energy Coulomb excitation.

The properties of halo nuclei 
depend very sensitively on the binding energy 
and despite  the ever increasing
precision of microscopic approaches using
realistic NN forces it will not be possible, say,
to predict the binding energies of nuclei to a level of 
about 100~keV.
Thus halo nuclei ask for 
new approaches in terms of some effective 
low-energy constants.
Such a treatment was provided in Ch.\ 2 and an example
with the one-neutron halo nucleus $^{11}$Be was given.
With the radioactive beam facilities at RIKEN,
GSI and RIA  one will be able to study also neutron halo
nuclei for intermediate masses in the years to come. 
This is expected to
be relevant also for the astrophysical r-process.
It is a great challenge to extend the present approach
for one-nucleon halo nuclei to more complicated cases,
like two-neutron halo nuclei.

The treatment of the continuum is a general problem,
which becomes more and more urgent when the dripline is approached.
In the present proceedings we studied 
the transition from bound to unbound states
as a typical example.

Recent experiments in the field of Coulomb dissociation
and the Trojan-horse method are discussed. It is a rich and fruitful 
development.

Beautiful experiments have to be matched with good theoretical
developments and painstaking analysis.
While the Coulomb dissociation method relies essentially
only on QED, precise experiments can give,
in combination with a thorough theoretical analysis,
precise answers for the astrophysical S factors. 
In the Trojan-Horse Method
more phenomenological aspects enter, 
like optical model parameters and effective nuclear interactions.
This makes the interpretation of the experimental
results in terms of astrophysical S factors 
less precise. However, this is not a new aspect in nuclear physics.
In the interpretation of screening effects one relies on the accuracy of the 
energy dependence. This is certainly better fulfilled than 
the accuracy of the
absolute values.
 
\begin{theacknowledgments}
We wish to thank Prabir Banerjee, Carlos Bertulani, Kai Hencken, 
Heinigerd Rebel, Radhey Shyam
and Dirk Trautmann for their collaboration on various topics 
in this field.   
\end{theacknowledgments}



\bibliographystyle{aipproc}   


%

\end{document}